\documentclass[prl,twocolumn,showpacs,superscriptaddress]{revtex4}

\usepackage{amsmath,amssymb,graphicx,color,bm}

\begin{document}

\title{Interactive optomechanical coupling with nonlinear polaritonic systems}

\author{N. Bobrovska}
\affiliation{Institute of Physics, Polish Academy of Sciences, Aleja Lotnikow 32/46, PL-02668 Warsaw, Poland}

\author{M. Matuszewski}
\affiliation{Institute of Physics, Polish Academy of Sciences, Aleja Lotnikow 32/46, PL-02668 Warsaw, Poland}

\author{T. C. H. Liew}
\affiliation{Division of Physics and Applied Physics, School of Physical and Mathematical Sciences, Nanyang Technological University, 21 Nanyang Link, Singapore 637371}

\author{O. Kyriienko}
\affiliation{The Niels Bohr Institute, University of Copenhagen, Blegdamsvej 17, DK-2100 Copenhagen, Denmark}

\date{\today}

\begin{abstract}
We study a system of interacting matter quasiparticles strongly coupled to photons inside an optomechanical cavity. The resulting normal modes of the system are represented by hybrid polaritonic quasiparticles, which acquire effective nonlinearity. Its strength is influenced by the presence of the mechanical mode and depends on the resonance frequency of the cavity. This leads to an {\it interactive} type of optomechanical coupling, being distinct from the previously studied dispersive and dissipative couplings in optomechanical systems. The emergent interactive coupling is shown to generate effective optical nonlinearity terms of high order, being quartic in the polariton number. We consider particular systems of exciton-polaritons and dipolaritons, and show that the induced effective optical nonlinearity due to the interactive coupling can exceed in magnitude the strength of Kerr nonlinear terms, such as those arising from polariton-polariton interactions. As applications, we show that the higher order terms give rise to localized bright flat top solitons, which may form spontaneously in polariton condensates.
\end{abstract}



\maketitle

{\it Introduction.---} In cavity optomechanics the optical properties of photonic systems are influenced by the motion of a mechanical oscillator \cite{AspelmeyerRev}. The coupling between optics and mechanics was shown to enable cooling of mechanical resonators \cite{Schliesser2006,Arcizet2006,Gigan2006,Usami2012}, thus allowing to test their quantum nature at the macroscale. Other examples, where merging of the concepts of optics and mechanics were shown to be useful are optical and mechanical squeezing \cite{Purdy2013,Safavi-Naeini2013,Wollman2015}, the generation of entanglement \cite{Palomaki2013,Riedinger2016}, gravitational wave detection \cite{LIGO2016}, optomechanically induced transparency (OMIT) \cite{Weis,Safavi-Naeini}, and quantum state teleportation \cite{Hou2016}. The related optical manipulation of mechanical degrees of freedom extensively contributed to the fields of precise measurements and sensing \cite{Bagci2014,Thompson2008,OConnel2010,Wilson2015,Lecocq2015}, many-body physics \cite{Ludwig2013,Xuereb2014,Peano2016}, and was theorized to have implications for quantum computing \cite{Hartmann2013} and communication \cite{Lukin}.

The essence of the optomechanical coupling is the dependence of an optical cavity parameter (say $\xi$) on the mechanical position [$\xi = \xi(x)$]. This can be achieved in several different ways. The common \textit{dispersive} coupling \cite{AspelmeyerRev} arises from the modulation of the frequency of the cavity mode, $\xi(x)= \omega_{\mathrm{C}}(x)$. A different coupling appears when a cavity decay is affected, leading to the \textit{dissipative} coupling, $\xi(x)=\gamma_{\mathrm{cav}}(x)$ \cite{Elste2009,Li2009,Kyriienko2014}. Simultaneously, these types of coupling can be classified to have leading \textit{linear} [$\xi(x) \propto x$] or \textit{quadratic} [$\xi(x) \propto x^2$] terms with respect to the position of an oscillator. Each type of coupling has distinct consequences for the properties and capabilities of a composite optomechanical system, introducing new physical phenomena.

In an apparently separate line of research, hybrid light-matter physics is studied using systems of semiconductor microcavities containing quantum wells. A wide variety of nonlinear effects have also been demonstrated~\cite{Carusotto2013}, including Bose-Einstein condensation~\cite{Kasprzak2006} and superfluidity~\cite{Amo2009,Sanvitto2011}. In these systems, direct excitons inside individual quantum wells are coupled to an optical mode and impart the effect of their Coulomb interaction as an effective Kerr nonlinearity on the resulting polariton modes. An interesting mechanism to enhance this nonlinearity has been based on using coupled quantum wells, supporting indirect-excitons in which electrons and holes in different quantum wells create excitons with large dipole moment. Alone, indirect-excitons can themselves form Bose-Einstein condensates~\cite{High2012}, but they can also be further coupled to an optical mode in a microcavity~\cite{Cristofolini2012}. The resulting dipolariton resonances can be continuously tuned, via an applied electric field, mixing the properties of indirect-excitons with direct-excitons and light~\cite{Wilkes2016,Sivalertporn2015,Byrnes2014,Rosenberg2016,KyriienkoLiew2014}.

Here, we consider a new form of optomechanical coupling, which arises from the dependence of the Kerr nonlinearity on the mechanical oscillator position, $\xi(x)= U(x)$. Appearing from the effective particle-particle interactions, this \textit{interactive} coupling contains high order terms, being fourth order in the particle number. This corresponds to a seventh order nonlinear susceptibility, typically inaccessible in optical systems~\cite{KivsharAgrawal_OpticalSolitons}, and introduces a highly nonlinear optical response.

As a potential platform for the realization of the interactive coupling we propose the hybrid optomechanical semiconductor system where a cavity mode is strongly coupled to an excitonic mode in a quantum well \cite{Kyriienko2014,Sete2015,Sete2012,Sarma2016,Fainstein2013,Anguiano2016}, forming polaritonic modes. There, it appears due to the position dependence of the excitonic fraction (Hopfield coefficient) of polaritons. In particular, we are interested in the system of dipolaritons, where high sensitivity of the Hopfield coefficients to small mechanical motion allows to generate high order nonlinear terms. Coming from the interactive coupling, they can become the dominant terms in the system, exceeding in strength the typical non-mechanical contribution to the Kerr type nonlinearity. We show that this very unique form of nonlinearity supports bright flat top solitons~\cite{Akhmediev_FLatTopSolitons}, which may form spontaneously in polariton condensates, representing a translational symmetry breaking.

{\it General model.---}We begin with a generic model which introduces the interactive optomechanical coupling, represented by a nonlinear optical cavity coupled to a mechanical oscillator. It can be described by the Hamiltonian $\hat{\mathcal{H}} = E_{\mathrm{C}} \hat{a}_{\mathrm{C}}^\dagger \hat{a}_{\mathrm{C}} + p^2/2m + m\omega_m^2 x^2/2 + U(x) \hat{a}_{\mathrm{C}}^\dagger \hat{a}_{\mathrm{C}}^\dagger \hat{a}_{\mathrm{C}} \hat{a}_{\mathrm{C}}$, where $E_{\mathrm{C}}$ denotes the energy of the cavity mode described by the creation (annihilation) operator $\hat{a}_{\mathrm{C}}^\dagger$ ($\hat{a}_{\mathrm{C}}$). The mechanical oscillator of mass $m$ and frequency $\omega_m$ is described by the momentum and position variables $p$ and $x$. $U(x)$ is a position dependent Kerr nonlinearity. The Hamiltonian can be expanded into series in $x$. Following the introduction of a phonon mode (mirror oscillation quanta) it can be written as
\begin{equation}
\label{eq:H_gen}
\hat{\mathcal{H}} = E_{\mathrm{C}} \hat{a}_{\mathrm{C}}^\dagger \hat{a}_{\mathrm{C}} + E_m \hat{b}^\dagger \hat{b} + U \hat{a}_{\mathrm{C}}^\dagger \hat{a}_{\mathrm{C}}^\dagger \hat{a}_{\mathrm{C}} \hat{a}_{\mathrm{C}} - g_i \hat{a}_{\mathrm{C}}^\dagger \hat{a}_{\mathrm{C}}^\dagger \hat{a}_{\mathrm{C}} \hat{a}_{\mathrm{C}} (\hat{b} + \hat{b}^\dagger),
\end{equation}
where $\hat{b}^\dagger$ ($\hat{b}$) is the creation (annihilation) operator for phonons. The position operator is defined as $\hat{x} = -x_{\mathrm{ZPF}} (\hat{b} + \hat{b}^\dagger)$, where $x_{\mathrm{ZPF}}$ is the zero point fluctuation of the mechanical oscillator \cite{AspelmeyerRev}. Here, $g_i = -x_{\mathrm{ZPF}} \partial U(x)/\partial x$ denotes the strength of the interactive coupling. Notably, the introduced form of optomechnical coupling is highly nonlinear, corresponding to the interaction being quartic in photon operators. The generation of large $g_i$ relies on finding a system where the effective optical nonlinearity has a steep position dependence.

{\it Polariton system.---}As a physical example of the system where interactive coupling appears naturally we describe the system of exciton-polaritons \cite{KavokinBook}. This arises from the strong coupling Hamiltonian
\begin{align}
\notag
\hat{\mathcal{H}}_{\mathrm{pol}} = &E_{\mathrm{C}}(x) \hat{a}_{\mathrm{C}}^\dagger \hat{a}_{\mathrm{C}} + E_{\mathrm{DX}} \hat{a}_{\mathrm{DX}}^\dagger \hat{a}_{\mathrm{DX}} + \frac{\Omega}{2} (\hat{a}_{\mathrm{C}}^\dagger \hat{a}_{\mathrm{DX}} + \hat{a}_{\mathrm{DX}}^\dagger \hat{a}_{\mathrm{C}}) \\
\label{eq:H_pol}  &+ U_{\mathrm{DX}} \hat{a}_{\mathrm{DX}}^\dagger \hat{a}_{\mathrm{DX}}^\dagger \hat{a}_{\mathrm{DX}} \hat{a}_{\mathrm{DX}},
\end{align}
where $\hat{a}_{\mathrm{DX}}^\dagger$ ($\hat{a}_{\mathrm{DX}}$) corresponds to an exciton creation (annihilation) operator with energy $E_{\mathrm{DX}}$. Here the index $_{\mathrm{DX}}$ corresponds to the direct exciton type formed by an electron and a hole located in the same quantum well. $\Omega$ corresponds to the exciton-photon coupling constant, typically referred as Rabi frequency. $U_{\mathrm{DX}}$ is the exciton-exciton interaction strength in the s-scattering limit, which typically arises from the electron and hole exchange processes. In the strong exciton-photon coupling regime, where the Rabi frequency is much larger than the decay rates in the system, $\Omega \gg \gamma_{\mathrm{C,DX}}$, the system can be rewritten in the polariton basis formed by superposition of cavity photon and exciton \cite{Kyriienko2014}. Considering only the lower energy mode, we obtain $\hat{\mathcal{H}}_{\mathrm{L}} = E_{\mathrm{L}}(x) \hat{a}_{\mathrm{L}}^\dagger \hat{a}_{\mathrm{L}} + U_{\mathrm{L}}(x) \hat{a}_{\mathrm{L}}^\dagger \hat{a}_{\mathrm{L}}^\dagger \hat{a}_{\mathrm{L}} \hat{a}_{\mathrm{L}}$,
%
%
where $\hat{a}_{\mathrm{L}}^\dagger$ ($\hat{a}_{\mathrm{L}}$) is the creation (annihilation) operator for the lower polariton mode. The important consequence of the strong coupling is the emergence of position dependence for the effective nonlinear term,
\begin{equation}
\label{eq:U_x}
U_{\mathrm{L}}(x) = |\beta(x) |^4 U_{\mathrm{DX}} = \frac{U_{\mathrm{DX}}}{4}\Big( 1 + \frac{(E_{\mathrm{C}}(x) - E_{\mathrm{DX}})}{\sqrt{(E_{\mathrm{C}}(x) - E_{\mathrm{DX}})^2 + \Omega^2}} \Big)^2,
\end{equation}
where $\beta(x)$ denotes a position dependent Hopfield coefficient, corresponding to the excitonic fraction of the lower polariton. Additionally, the energy of the lower polariton contains a dispersive interaction inherited from the cavity mode, $E_{\mathrm{L}}(x)= (E_{\mathrm{C}}(x)+E_{\mathrm{DX}})/2 - \sqrt{(E_{\mathrm{C}}(x) - E_{\mathrm{DX}})^2 + \Omega^2}/2$. The expansion of Eq. (\ref{eq:U_x}) reveals the position-independent Kerr term for the lower polariton mode, $U = |\beta(0) |^4 U_{\mathrm{DX}}$, and the interactive coupling term given by $g_{i,\mathrm{L}} = - x_{\mathrm{ZPF}} U_{\mathrm{DX}} \partial |\beta(x)|^4/\partial x$. This ultimately relies on the exchange based direct exciton interaction and the derivative of the Hopfield coefficient, which in the case of polaritons has a shape governed by the frequency detuning. This largely limits the tunability of the coupling. In order to introduce an in-situ controllable coupling and enlarge the magnitude of the interactive coupling we consider a more general dipolaritonic setup where indirect excitons are also present.

{\it Dipolariton system.---} We begin with the Hamiltonian describing the optomechanical coupling in a dipolariton microcavity \cite{Cristofolini2012}:
\begin{align}
&\hat{\mathcal{H}}_{\mathrm{dpl}}=E_{\mathrm{C}}(x)\hat{a}_{\mathrm{C}}^\dagger\hat{a}_{\mathrm{C}} + E_{\mathrm{DX}}\hat{a}_{\mathrm{DX}}^\dagger\hat{a}_{\mathrm{DX}} + E_{\mathrm{IX}}\hat{a}_{\mathrm{IX}}^\dagger\hat{a}_{\mathrm{IX}} +\hbar\omega_m\hat{b}^\dagger\hat{b} \notag\\
&+\left(\frac{\Omega}{2}\hat{a}_{\mathrm{DX}}^\dagger\hat{a}_{\mathrm{C}} + \frac{J}{2} \hat{a}_{\mathrm{IX}}^\dagger\hat{a}_{\mathrm{DX}} + \mathrm{h.c.}\right) +U_{\mathrm{DX}}\hat{a}_{\mathrm{DX}}^\dagger\hat{a}_{\mathrm{DX}}^\dagger\hat{a}_{\mathrm{DX}}\hat{a}_{\mathrm{DX}} \notag\\
&+ U_{\mathrm{IX}}\hat{a}_{\mathrm{IX}}^\dagger\hat{a}_{\mathrm{IX}}^\dagger\hat{a}_{\mathrm{IX}}\hat{a}_{\mathrm{IX}} +U_{\mathrm{DI}}\hat{a}_{\mathrm{DX}}^\dagger\hat{a}_{\mathrm{IX}}^\dagger\hat{a}_{\mathrm{DX}}\hat{a}_{\mathrm{IX}},
\end{align}
where, together with cavity photons and direct excitons, indirect excitons---spatially separated electron-hole complexes \cite{Butov2007,Snoke2013}---are present. Here $\hat{a}_{\mathrm{IX}}^\dagger$ ($\hat{a}_{\mathrm{IX}}$) is the creation (annihilation) operator of the indirect exciton mode with energy $E_{\mathrm{IX}}$. $J$ is the tunnelling coupling between direct exciton and indirect exciton. $U_{\mathrm{IX}}$, $U_{\mathrm{DX}}$ and $U_{\mathrm{DI}}$ characterize the strength of repulsive interactions between pairs of direct excitons, pairs of indirect excitons, and mixed direct-indirect exciton pairs.

\begin{figure}[h!]
\includegraphics[width=\columnwidth]{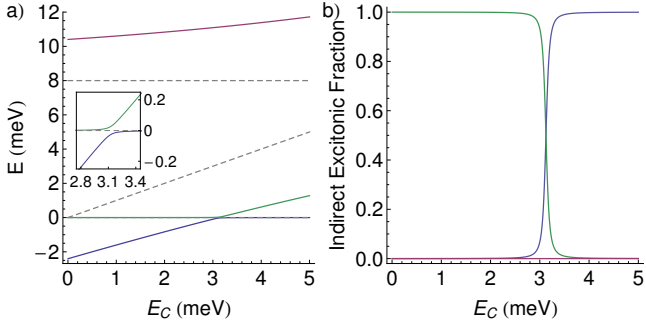}
\caption{(color online) (a) Energy of dipolaritonic modes. Dashed lines show the uncoupled cavity photon and exciton energies. (b) Indirect exciton fraction (Hopfield coefficient) for the three polariton modes. Parameters: $J=0.1$ meV, $\Omega=5$ meV, $E_{\mathrm{DX}}=8$ meV, $E_{\mathrm{IX}}=0$.}
\label{fig:Hopfield}
\end{figure}

Diagonalization of the linear part of the Hamiltonian yields three dipolariton branches, shown in Fig.~\ref{fig:Hopfield}(a) for typical parameters.
Here we have chosen $\Omega\gg J$. For differences in the bare energies, $E_{\mathrm{C}}$, $E_{\mathrm{DX}}$ and $E_{\mathrm{IX}}$ comparable to $\Omega$, we ensure that all dipolariton modes have a significant photonic fraction. The smaller value of $J$ also admits a narrow anticrossing of the direct and indirect exciton modes. Considering the behaviour of the lower dipolariton mode, this results in a rapid variation of the indirect exciton fraction with cavity mode energy as shown in Fig.~\ref{fig:Hopfield}(b). This rapid variation allows the system to be sensitive to small changes in the mirror position.

To describe the interactive optomechanical coupling in the dipolariton system, we switch to the diagonal basis and consider only the lower mode, with the Hamiltonian:
\begin{align}
\hat{\mathcal{H}}_{\mathrm{L}}=& E_{\mathrm{L}}(x)\hat{\psi}^\dagger\hat{\psi}+\hbar\omega_m\hat{b}^\dagger\hat{b} +\big[U_{\mathrm{DX}}\beta_\mathrm{D}(x)^4 \notag\\ &+U_{\mathrm{IX}}\beta_\mathrm{I}(x)^4 + U_{\mathrm{DI}}\beta_\mathrm{D}(x)^2\beta_\mathrm{I}(x)^2\big]\hat{\psi}^\dagger\hat{\psi}^\dagger\hat{\psi}\hat{\psi},
\label{eq:HamLP}
\end{align}
where $\hat{\psi}$ denotes the field operator for lower dipolaritons. $\beta_\mathrm{D}(x)$ and $\beta_\mathrm{I}(x)$ are the (Hopfield) factors representing the direct and indirect exciton fractions of the lower dipolariton state. Since $U_{\mathrm{IX}}>U_{\mathrm{DI}}>U_{\mathrm{DX}}$ the term in square brackets is particularly sensitive to $x$.

The $x$-dependent terms in Hamiltonian~(\ref{eq:HamLP}) can be expanded about the point $x=0$, which we take as corresponding to the anticrossing point of direct and indirect excitons. The lower polariton energy then reads
\begin{align}
E_\mathrm{L}(x)\approx E_\mathrm{L}(0)+x\frac{\partial E_{\mathrm{L}}}{\partial x}=E_\mathrm{L}(0)-g_0\left(\hat{b}+\hat{b}^\dagger\right)\frac{\partial E_\mathrm{L}}{\partial E_\mathrm{C}},
\end{align}
where we replaced $x$ with its quantum description $\hat{x}=x_\mathrm{ZPF}(\hat{b}+\hat{b}^\dagger)$ and defined the constant $g_0=-x_\mathrm{ZPF}\partial E_\mathrm{C}/\partial x$. Similarly we expand Hopfield coefficients as
\begin{align}
\label{eq:beta_4}
&\beta_{\mathrm{D,I}}(x)^4 \approx \beta_{\mathrm{D,I}}(0)^4 - g_0\left(\hat{b}+\hat{b}^\dagger\right)\frac{\partial\beta_{\mathrm{D,I}}^4}{\partial E_\mathrm{C}},\\
\label{eq:beta_2_2}
&\beta_\mathrm{D}(x)^2\beta_\mathrm{I}(x)^2 \approx \beta_\mathrm{D}(0)^2\beta_\mathrm{I}(0)^2 - g_0\left(\hat{b}+\hat{b}^\dagger\right)\frac{\partial(\beta_\mathrm{D}^2\beta_\mathrm{I}^2)}{\partial E_\mathrm{C}}.
\end{align}

The polariton-mechanical Hamiltonian~(\ref{eq:HamLP}) becomes:
\begin{align}
\hat{\mathcal{H}}_{\mathrm{L}}&=E_\mathrm{L}(0)\hat{\psi}^\dagger\hat{\psi}+\hbar\omega_m\hat{b}^\dagger\hat{b} + [U_{\mathrm{DX}}\beta_\mathrm{D}(0)^4 + U_{\mathrm{IX}}\beta_\mathrm{I}(0)^4 \notag\\
& + U_{\mathrm{DI}}\beta_\mathrm{D}(0)^2\beta_\mathrm{I}(0)^2]\hat{\psi}^\dagger\hat{\psi}^\dagger\hat{\psi}\hat{\psi} + \hat{\zeta} \left(\hat{b}+\hat{b}^\dagger\right),
\label{eq:H_L_fin}
\end{align}
where we defined the operator
\begin{align}
\hat{\zeta}=&-g_0\left(U_{\mathrm{DX}}\frac{\partial \beta_\mathrm{D}^4}{\partial E_\mathrm{C}}+U_{\mathrm{DI}}\frac{\partial(\beta_\mathrm{D}^2\beta_\mathrm{I}^2)}{\partial E_\mathrm{C}}
+U_{\mathrm{IX}}\frac{\partial \beta_\mathrm{I}^4}{\partial E_\mathrm{C}}\right)\hat{\psi}^\dagger\hat{\psi}^\dagger\hat{\psi}\hat{\psi}\notag\\
&-g_0\frac{\partial E_\mathrm{L}}{\partial E_\mathrm{C}}\hat{\psi}^\dagger\hat{\psi}.
\label{eq:zeta}
\end{align}
The last term in Hamiltonian (\ref{eq:H_L_fin}) represents the interactive coupling for the dipolariton mode.

{\it Effective Hamiltonian.---}We now seek to eliminate the phonon modes using the polaron (Schrieffer-Wolff) transformation. We consider the unitary transformation $\hat{\mathcal{H}}_{\mathrm{L}}^\prime=e^{\hat{S}}\hat{\mathcal{H}}e^{-\hat{S}}$, where $\hat{S}=\hat{\zeta}\left(\hat{b}^\dagger-\hat{b}\right)/\hbar\omega_m$ \cite{Rabl2011}. The transformed Hamiltonian can be rewritten using the Baker-Campbell-Hausdorff formula, $e^{\hat{S}}\hat{\mathcal{H}}e^{-\hat{S}}=\sum_{m=0}^\infty\frac{1}{m!}[\hat{S},\hat{\mathcal{H}}]_m$, where $[\hat{A},\hat{B}]_m$ denotes the nested commutator of order $m$. Given that the commutator $\left[\hat{\psi}^\dagger\hat{\psi},\hat{\psi}^\dagger\hat{\psi}^\dagger\hat{\psi}\hat{\psi}\right]=0$, only the first three terms of the sum are non-zero. Then
\begin{align}
&\hat{\mathcal{H}}_{\mathrm{L}}^\prime = E_\mathrm{L}(0)\hat{\psi}^\dagger\hat{\psi}+\hbar\omega_m\hat{b}^\dagger\hat{b} + [U_{\mathrm{DX}}\beta_\mathrm{D}(0)^4 + U_{\mathrm{IX}}\beta_\mathrm{I}(0)^4 \notag\\
&+ U_{\mathrm{DI}}\beta_\mathrm{D}(0)^2\beta_\mathrm{I}(0)^2]\hat{\psi}^\dagger\hat{\psi}^\dagger\hat{\psi}\hat{\psi} -\frac{\hat{\zeta}^2}{\hbar\omega_m},
\label{eq:Heff}
\end{align}
where we note that the last term contains interactive contributions up to eighth order in the field operators.

{\it Parameter Estimates.---}From the above Hamiltonian we see that the optomechanical interactive coupling has a typical strength characterized by the energy shift
\begin{equation}
\Delta=-\frac{\left(g_0 U_{\mathrm{IX}}\partial \beta_{\mathrm{I}}^4/\partial E_{\mathrm{C}}\right)^2}{\hbar\omega_m}N^3,
\end{equation}
where $N$ is the average dipolariton number and for an order of magnitude estimate we will focus on the indirect exciton interactions characterized by $U_{\mathrm{IX}}$. The above term competes with the usual repulsive interactions, leading to the typical blueshift $\Delta_0=U_{\mathrm{IX}} N \beta_{\mathrm{I}}(0)^4$.

To present realistic estimates for the system, we infer the parameters from the optomechanical sample studied in Ref. \cite{Fainstein2013}. Considering a quasi-1D structure of length $L = 200~\mu$m and width $d = 1~\mu$m, the optomechanical coupling strength $g_0$ can be estimated with conservative parameters $\omega_m=2\pi\times 1.25$~GHz, $x_{\mathrm{ZPF}}=1.5 \times10^{-7}$~nm, $\hbar^{-1}\partial E_\mathrm{C}/\partial x=2\pi\times100$~THz/nm, which gives $g_0 \approx 6.2\times10^{-5}$~meV. The slope of the Hopfield coefficient can be estimated as $\partial \beta_\mathrm{I}^4/\partial E_\mathrm{C} \sim 1/J$, and we take $J=0.05$~meV with $\beta_\mathrm{I}^2(0)=1/2$. The indirect exciton interaction corresponds to $U_\mathrm{IX}=10^{-4}$~meV \cite{Kyriienko2012}, with normalization area of $A=200~\mu$m$^2$ and 6~nm separation between quantum wells of 12~nm width. The ratio between the energy shifts then reads $|\Delta|/|\Delta_0| \approx 1.2 \times 10^{-7} N^2$. For a polariton density $N/A \sim 14~\mu$m$^{-2}= 1.4\times10^{9}$~cm$^{-2}$, we can then expect a cross-over into an interactive coupling regime. Furthermore, in Ref.~\cite{Jusserand2015} very high optomechanical coupling strengths of $g_0=0.1\times2\pi$ THz have been suggested, which would allow further enhancement of the interactive coupling.

{\it Spatial Dynamics.---} The spatial dynamics of a polaritonic system at large occupation numbers can be typically treated within the mean-field approximation~\cite{Carusotto2013}. This corresponds to the description of the system in terms of a mean-field wavefunction, $\hat{\psi} \rightarrow \psi$. Here we consider the behavior of optomechanical dipolaritons in a quasi-one-dimensional system, which could be implemented experimentally by etching one-dimensional channels~\cite{Wertz2012,Gao2012}. From the effective Hamiltonian [Eq.~(\ref{eq:Heff})], we derive a Gross-Pitaevskii-type equation for the 1D polariton wavefunction, $\psi(z,t)$:
\begin{align}
\label{eq:GP1D}
i\hbar&\frac{\partial\psi(z,t)}{\partial t}=\left[-\frac{\hbar^2\nabla^2}{2m_\mathrm{LP}}+\left(\alpha_\mathrm{1D}-\frac{\gamma_\mathrm{1D}^2}{\hbar\omega_m}\right) n\right.\\
&\left.-\frac{2\beta_\mathrm{1D}\gamma_\mathrm{1D} n^2+\beta_\mathrm{1D}^2n^3}{\hbar\omega_m}+\delta n^4\right]\psi(z,t),\notag
\end{align}
where $z$ denotes a coordinate along the waveguide, and we define $n=|\psi(z,t)|^2$. Here we introduced the dispersion of polaritons, characterized by an effective mass $m_\mathrm{LP}$. We consider a high Q-factor microcavity, approaching the equilibrium limit~\cite{Snoke_LongLifetimeEquilibrium,Nelsen2013}, and consequently we neglected the nonequilibrium terms. The nonlinear interaction parameters now have dimensions consistent with a 1D polariton density $n$, and are related to previously described parameters by $\alpha_\mathrm{1D}\approx U_{\mathrm{IX}} A \beta_\mathrm{I}(0)^4/d$, $\beta_\mathrm{1D}\approx g_0 U_{\mathrm{IX}}/J(A/d)^{3/2}$, and $\gamma_\mathrm{1D}=g_0(A/d)^{1/2} \partial E_\mathrm{L}/\partial E_\mathrm{C}\approx g_0(A/d)^{1/2}/2$. 

\begin{figure}[h!]
\includegraphics[width=\columnwidth]{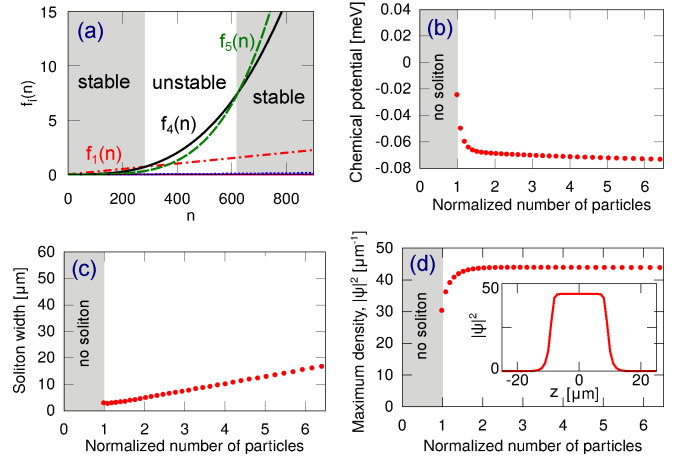}
\caption{(color online) (a) Comparison of the term amplitudes containing $n$ in Eq.~\ref{eq:GP1D}. $f_1(n)=\alpha n$, $f_2(n)=\frac{\gamma_\mathrm{1D}^2}{\hbar\omega_m} n$, $f_3(n)=\frac{2\beta_\mathrm{1D}\gamma_\mathrm{1D}}{\hbar\omega_m}n^2$, $f_4(n)=\frac{\beta_\mathrm{1D}^2}{\hbar\omega_m}n^3$, $f_5(n)=\delta n^4$. The other panels show the dependence of the chemical potential (b), soliton width (c), and soliton height (d) with respect to normalized number of particles $N/N_0$ where $N_0=119$. The inset in panel (d) shows the flat-top  soliton profile at $N/N_0 = 6$. Parameters: $\alpha_\mathrm{1D}=0.005$~meV$\mu$m, $\beta_\mathrm{1D}=3.54\times10^{-4}$~meV$\mu$m$^{3/2}$, $\gamma_\mathrm{1D}=4.42\times10^{-4}$~meV$\mu$m$^{1/2}$, $\delta=0.5\times 10^{-6}$~meV$\mu$m$^4$; $m_\mathrm{LP}$ was taken as $1.5\times10^{-4}$ of the free electron mass.}
\label{fig:functions}
\end{figure}

To prevent the unphysical collapse of the condensate wavefunction to an infinitesimally small point, which would be caused by the high order attractive nonlinearity, we phenomenologically account for the higher order repulsive interactions by introducing the term $\delta n^4$ in Eq.~(\ref{eq:GP1D}) \cite{Berge1998}. The small parameter $\delta>0$ allows for the stabilization of the condensate. The relative strength of the different  terms in Eq.~(\ref{eq:GP1D}) is compared in Fig.~\ref{fig:functions}(a). The attractive high order nonlinearity, in particular the $f_4(n)$ term, can be expected to cause the condensate to compress and increase its density up to the point where the repulsive $\delta n^4$ term becomes dominant.

Using Eq.~(\ref{eq:GP1D}) and the imaginary time method, we investigate the formation of a condensate in the ground state of the interacting system. By propagating in imaginary time, we effectively probe the relaxation of condensate energy and the formation of a state corresponding to an energy minimum.
\begin{figure}[h]
\centering
\includegraphics[width=1.0\linewidth]{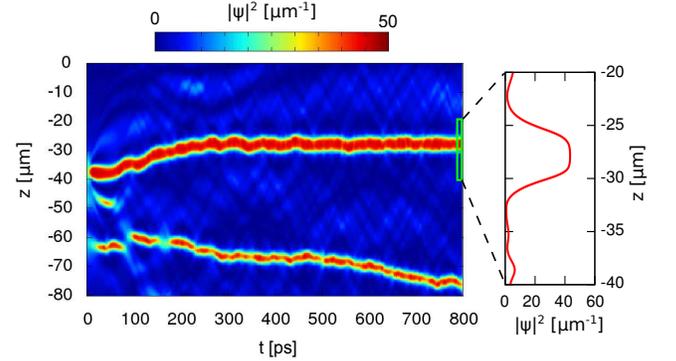}
\caption{(color online) Stable flat-top solitons, emerging from an initial random noise, obtained by propagating Eq.~(\ref{eq:GP1D}) in real time. Parameters were the same as in Fig.~\ref{fig:functions}.}
\label{fig:solitons}
\end{figure}
We find that when the number of particles exceeds a certain threshold, the ground state takes the form of a flat-top soliton. In Fig. {\ref{fig:functions}}(b)-(d) we show the variation of the calculated chemical potential, soliton width and soliton height as a function of the total number of particles in the system, $N$. With increasing total particle number the solitons have an almost fixed height (in intensity) while their spatial size increases approximately linearly.

The simulation in Fig.~\ref{fig:solitons} shows an example of spontaneously formed solitons in an out-of-equilibrium situation. Starting from a random initial condition and propagating in real time, three flat-top solitons are spontaneously formed. While the final state is not the ground state, excess energy is transformed into small excitations. This shows that the flat-top solitons are attractors of the evolution even in the absence of relaxation, which is a property of dynamically stable solitons.

{\it Conclusions.---}We have introduced the notion of interactive optomechanical coupling, which appears in generic systems where the strength of Kerr nonlinearity is influenced by a mechanical motion. It has a highly nonlinear character, and manifests itself as a seventh order nonlinear susceptibility. As particular examples of systems where interactive coupling can be realized we describe exciton-polaritons and dipolaritons. In the latter case the steep dependence of Hopfield coefficients allows to attain large interactive coupling constant. The theory was applied to 1D polaritonic waveguides, revealing the generation of bright flat-top solitons in polaritonic fluids.

{\it Acknowledgements.---}O.K. thanks I. A. Shelykh and A. S. S{\o}rensen for the useful discussions, and acknowledges the support from the European Union Seventh Framework Programme through the ERC Grant QIOS (Grant No. 306576). NB and MM acknowledge support from National Science Center, Poland grants DEC-2011/01/D/ST3/00482 and 2015/17/B/ST3/02273. T.L. ackoweledges support from the MOE Academic Research Fund (2016-T1-001-084).

\end{document}